\documentclass[preprints,article,submit,pdftex,moreauthors]{Definitions/mdpi} 
\firstpage{1} 
\pubvolume{12}
\issuenum{1}
\articlenumber{1620}
\pubyear{2022}
\copyrightyear{2022}
\externaleditor{Academic Editors: Claudia Pacelli, Francesca Ferranti and Marta del Bianco} 
\datereceived{20 September 2022} 
\dateaccepted{14 October 2022} 
\datepublished{17 October 2022} 
\hreflink{https://doi.org/10.3390/life12101620} 

\Title{On the Role of $^{40}$K in the Origin of Terrestrial Life}

\TitleCitation{On the Role of $^{40}$K in the Origin of Terrestrial Life}

\Author{Giovanni Vladilo \orcidA{}}

\AuthorNames{Giovanni Vladilo}

\AuthorCitation{Vladilo, G.}

\address[1]{%
 INAF-Osservatorio Astronomico di Trieste, Via G.B. Tiepolo 11, 34143 Trieste, Italy; giovanni.vladilo@inaf.it \mbox{Tel.: +39-0403199216}}

\abstract{The abundance and biological role of potassium suggest that its unstable nuclide was present in all stages of terrestrial biogenesis. With its enhanced isotopic ratio in the Archean eon, $^{40}$K may have contributed to the special, perhaps unique, biogenetic conditions that were present in the primitive Earth. 
Compared to the U and Th radionuclides, $^{40}$K has a less disruptive radiochemical impact, which
may drive a moderate, but persistent evolution of the structural and functional properties of proto-biological molecules.
In the main $\beta$-decay route of $^{40}$K, the radiation dose generated by an Archean solution with potassium ions can be larger than the present background radiation on Earth by one to two orders of magnitude. Estimates of the rates of organic molecules indirectly affected by $\beta$ decays are provided for two schematic models of the propagation of secondary events in the solvent of prebiotic solutions. The left-handed $\beta^-$ particles emitted by $^{40}$K are the best candidates to trigger an enantiomeric excess of L-type amino acids via weak nuclear forces in the primitive Earth. The concentration-dependent radiation dose of $^{40}$K fits well in dry--wet scenarios of life's origins and should be considered in realistic simulations of prebiotic chemical pathways.
}


\begin{document}

%
%

\section{Introduction}

In the most-common scenario of life's origins, initially proposed by Oparin one century ago \cite{opa24},
terrestrial life emerges as a result of natural processes driving an increase in molecular complexity and functionality.
However, despite remarkable progress in proposing plausible prebiotic chemical pathways \cite{lui16,sal19},
the reconstruction of the sequence of physico-chemical processes that led to the emergence of life 
in different environments of the primitive Earth \cite{bad02,mul12,wes18,ton19,dam20}
remains one of the most challenging problems of science.
Among the physical processes considered in studies of the origins of life,
here, we focus on the radiochemical effects generated by natural sources of radiation. 

\textls[-5]{Throughout the history of the Earth, radionuclides of astrophysical origin captured by our planet 
have generated radiogenic heating and triggered radiation chemistry \mbox{reactions \cite{dra93}.} }
Radiogenic heating of the Earth's mantle facilitates plate tectonics \cite{and07},
which is believed to support the long-term habitability of our planet \cite{kas93,unt22}. 
Radiation chemistry is well known for its biological effects \cite{dra93,cho02,mag05},
but only in the last two decades has it been investigated as a potential trigger of prebiotic chemical reactions \cite{gar01,zag03,par04,ada07,dar11,alt20}. 
Indeed, natural radioactivity affects the bonds in organic molecules, 
which could be broken and reformed to build larger molecules, ancestors of extant biopolymers.
Furthermore, the radiolysis of water can be a starting point for a chain of chemical reactions of prebiotic interest.
Most studies of natural radioactivity in the prebiotic scenario have considered the radionuclides of uranium and thorium \cite{par04,ada07,alt20},
which provide most of the internal heating of the Earth's mantle. 
In this work, we instead investigate the potential role of a lighter radionuclide, $^{40}$K,
which has been rarely and only marginally discussed in previous studies of life's \mbox{origins \cite{noy77,dra91,zag03}.}
The general purpose of this paper is to investigate the potential role of $^{40}$K in the stages 
of prebiotic molecular evolution when early polymers were starting to emerge. 
Two specific points deal with the biogenic conditions of the Archean and the origin of \mbox{biomolecular chirality.}

The physico-chemical conditions of the Earth during the Archean
were different from those of the following eons in many respects
(e.g., higher rotation rate, lower fraction of continents, anoxic atmosphere, high rate of UV photons, fainter, but more active Sun, etc.). 
Since terrestrial life's originated in that epoch, it is important to understand
which Archean conditions might have been essential for the emergence of life.
A specific motivation of this study is to understand if the radiochemical effects of $^{40}$K, 
whose isotopic ratio was significantly enhanced in the Archean,
may have contributed to the special conditions that allowed life to emerge.

The origin of biomolecular homochirality is an open question in studies of life's \linebreak{origins \cite{ben12,lui16,bla19}.} 
According to some authors \cite{joy84,bla19}, the homochirality must have originated before the prebiotic
synthesis of monomers, since polymerization tends to be inhibited in racemic mixtures of nucleotides.
The general idea for the origin of molecular chirality is that, after an initial, small enantiomeric enhancement, 
some form of chirality \mbox{amplification \cite{yam66,ber99I,ber99II,bry20}}
allowed homochiral polymers to be assembled from enantiomerically pure building blocks. 
Different types of natural processes have been invoked to explain the generation of an initial enantiomeric excess \cite{bla19}.
Processes forcing a well-defined chirality (either left or right) are appealing in this context because a steady forcing in a constant direction would increase the chance of accumulating an enantiomeric excess. 
This is important in the prebiotic scenario,
where racemization converts optically active molecules back into a racemic mixture in
geologically short time scales \cite{bad82}.
Parity violation of electroweak forces \cite{bet14,mar17} is a natural process, which, in principle,
may provide a universal explanation of an initial enantiomeric excess \cite{lee56,wu57,kes77,sal91}. 
In the context of parity violation interpretations, here, we examine the possibility that the spin-polarized $\beta^-$ particles 
emitted by $^{40}$K may have generated a small enantiomeric excess of early polymers.
The possibility that spin-polarized electrons produced by UV irradiation of Archean magnetite deposits
may have triggered a prebiotic chiral excess has been considered in a recent paper \cite{fur22}.

The paper is structured as follows. 
The reasons why, at variance with heavier radionuclides, $^{40}$K was likely
to be present in all stages of prebiotic chemistry 
are discussed in Section \ref{theKcase}. 
Quantitative estimates of the radiochemical impact generated by $\beta^-$ decays $^{40}$K 
in a prebiotic solution of the primitive Earth are provided in Section \ref{40Kradiochemistry}.
The possibility that $^{40}$K can play a role in the origin of biomolecular chirality
is discussed in Section \ref{chirality}. 
The conclusions and suggestions for future experiments are summarized in Section \ref{conclusions}.
Basic properties of the $^{40}$K decay can be found in Appendix \ref{properties40K}. 
We assumed that life emerged at $\approx$4 Ga, i.e., 
in the time span between 4.3--4.2 Ga, the age of the formation of oceans \cite{pin06}, 
and 3.75--3.5 Ga, the age of the oldest, best-established traces of life \cite{wes06}. 

\section{Sources of Natural Radiation in the Early Archean: The Case for \boldmath{$^{40}$K}}
\label{theKcase} 

Several sources of natural radiation were present on Earth in the early Archean.
In principle, any of them may have influenced the prebiotic chemical evolution leading to
the emergence of the first functional molecules with catalytic--genetic properties.
To ascertain the relative importance of the different sources of radiation, we considered the following aspects:
(1) the abundance and distribution of the sources on the primitive Earth, 
(2) the potential role of the stable isotopes in prebiotic chemistry,
(3) the compatibility of the decay products with protobiological molecular structures,
(4) the continuity with present-day terrestrial life, and
(5) the enhancement of the radiation activity in the early Archean.
The properties of some natural radionuclides present on Earth are summarized in Table \ref{tabRadionuclides}. 
\begin{table}[H]
\caption{Properties of natural radionuclides in the Earth's crust and sea \cite{crc17,cam04}. 
\label{tabRadionuclides}}
\begin{adjustwidth}{-\extralength}{0cm}
\setlength{\cellWidtha}{\fulllength/10-2\tabcolsep-0in}
\setlength{\cellWidthb}{\fulllength/10-2\tabcolsep+0.3in}
\setlength{\cellWidthc}{\fulllength/10-2\tabcolsep-0in}
\setlength{\cellWidthd}{\fulllength/10-2\tabcolsep-0in}
\setlength{\cellWidthe}{\fulllength/10-2\tabcolsep-0in}
\setlength{\cellWidthf}{\fulllength/10-2\tabcolsep-0in}
\setlength{\cellWidthg}{\fulllength/10-2\tabcolsep-0in}
\setlength{\cellWidthh}{\fulllength/10-2\tabcolsep-0in}
\setlength{\cellWidthi}{\fulllength/10-2\tabcolsep-0.3in}
\setlength{\cellWidthj}{\fulllength/10-2\tabcolsep-0in}
\scalebox{1}[1]{\begin{tabularx}{\fulllength}{>{\centering\arraybackslash}m{\cellWidtha}>{\centering\arraybackslash}m{\cellWidthb}>{\centering\arraybackslash}m{\cellWidthc}>{\centering\arraybackslash}m{\cellWidthd}>{\centering\arraybackslash}m{\cellWidthe}>{\centering\arraybackslash}m{\cellWidthf}>{\centering\arraybackslash}m{\cellWidthg}>{\centering\arraybackslash}m{\cellWidthh}>{\centering\arraybackslash}m{\cellWidthi}>{\centering\arraybackslash}m{\cellWidthj}}
\toprule
\textbf{Elem.} & \textbf{Abund.} & \textbf{Abund.} & \textbf{Role in} &
\textbf{Unstable} & \textbf{Isotopic} & \textbf{Half} & \textbf{Decay} & \boldmath{$p_i$} & \textbf{Decay} \\
& \textbf{in the Crust } & \textbf{in the Sea} & \textbf{Biology} & \textbf{Nuclide} & \textbf{ Ratio} & \textbf{Life} & \textbf{Mode} & &
\textbf{Product} \\
& \textbf{(mg/kg)} & \textbf{(mg/L)} & & & \textbf{(\%) }&\textbf{ (Gyr)}& & \\
\midrule
K & 20900 & 399 & yes & $^{40}$K	 & 0.0117 	& $1.248$ & $\beta^-$ & 0.893 & $^{40}$Ca \\ 
 	& & &	& & & & EC,$\gamma$ & 0.107 & $^{40}$Ar \\ 
Th &9.6 & $1 \times 10^{-6}$ & no & $^{232}$Th & 100 	& $14 $ & $\alpha$ & & $^{228}$Ra \\
U & $2.7$ & $3.2 \times 10^{-3}$ & no & $^{235}$U		& 0.72 & $0.704$	& $\alpha$ & & $^{231}$Th \\ 
 & & & & $^{238}$U & 99.27		& $4.46$	& $\alpha$ & &$^{234}$Th \\ 
	\bottomrule
		\end{tabularx}}
	\end{adjustwidth}
\end{table}
\subsection{Abundance and Distribution of Natural Radiation Sources on the Primitive Earth}
 
Potassium is among the eight most-abundant elements in the Earth's crust and sea,
where it is several orders of magnitude more abundant 
than thorium or uranium 
\cite{crc17,are09}. While these heavier elements are concentrated in specific environments \cite{par04,ada07,alt20}, 
potassium is widespread and likely to be present in environments 
considered to be plausible sites for the emergence of terrestrial life, 
such as evaporative alkaline lakes \cite{ton19}
and subaerial hot spring fields \cite{dam20,djo21,kra21}.
The concentration of potassium was moderately higher in Archean oceans ($C_\text{K}$ = 14--24 mM \cite{har03}) 
than in present-day oceans ($C_\text{K}$ = 10.2 mM \cite{crc17}). 
Since it is abundant both in the crust and in the sea, potassium
and its unstable isotope $^{40}$K should be considered 
in all possible scenarios of life's origins, whether subaerial in early emerged lands
or in shallow waters 
\cite{wes18,mul12,ben12}.

\subsection{Potential Role of the Stable Potassium Isotope in Prebiotic Chemistry}

Laboratory studies have shown that potassium 
can be important in some stages of prebiotic chemistry.
Specifically, potassium has the potential to assist 
the formation of the first membranes \cite{nat10} 
and also the assemblage of peptides \cite{dub13}. 
Remarkably, the molar concentration found in many Archaea, $C_\text{K} \approx 1 $\,M \cite{mar99}, 
lies in a range that appears to be optimal for the assemblage of peptides \cite{dub13}. 

\subsection{Compatibility of Decay Products with Protobiological Molecular Structures}
\label{compatibility} 

Among the radionuclides listed in Table \ref{tabRadionuclides}, $^{40}$K is the only one that
has decay particles and decay products that can coexist with the early functional molecules and protocells. 
This conclusion, which is in line with the presence 
of $^{40}$K and the absence of heavy radionuclides in extant life,
{is based on the following arguments. 
%
The radionuclides of Th and U emit $\alpha$ particles with energies of $\simeq$4 MeV
and generate unstable nuclides (Table \ref{tabRadionuclides}), which in turn produce chains of decays.
In each of these chains, several $\alpha$ particles with \mbox{energies $> 4$\,MeV} are emitted. 
At variance with this behavior, $^{40}$K emits $\beta$ and $\gamma$ rays with \mbox{energies $\leq 1.5$\,MeV}
(Appendix \ref{properties40K})
and generates stable nuclides, such as $^{40}$Ca, 
which is used in terrestrial biochemistry, and $^{40}$Ar,
which might play some secondary biological role \cite{yez13}. 
Another remarkable difference is 
the linear energy transfer (LET), i.e., the mean energy transferred to the medium
per unit path length traveled by the ionizing particle.
Typical values of LET in water \cite{dra93} 
are three orders of magnitude larger for $\alpha$ particles ($\simeq$1.5 $\times 10^2$ keV/$\upmu$m)
than for $\beta$ particles ($\simeq$0.2 keV/$\mu$m); 
the LET of $\gamma$ rays is lower than that of $\beta$ particles (Appendix \ref{properties40K}).
These facts suggest that the strong activity of Th and U radionuclides 
may disrupt proto-biological molecular structures, 
whereas $^{40}$K can coexist with such structures, affecting their evolution in a non-disruptive way.}

\subsection{Continuity with Present-Day Terrestrial Life}

The role of potassium in present-day life makes this element special
compared to other elements that possess natural radioisotopes. 
While thorium and uranium do not have any biological role, 
potassium is an essential ingredient of ionic channels.
%
The present-day potassium channel is an archetype of other structures of ionic channels \cite{doy98,ber07},
suggesting that potassium was essential also in the earliest forms of life. 
In line with this possibility is the large concentration of intracellular potassium 
found in Archaea organisms, which lie close to the root of the phylogenetic tree \cite{woe77}. 
Indeed, the molar concentration (\mbox{M = mole/liter}) of potassium
($C_\text{K}$) in the cytosol of Archaea is generally above 0.5 M \cite{mar99}.
For Archaea growing in saline habitats, such as {\it Halobacterium halobium}, this high $C_\text{K}$ can be attributed to osmoadaptation. However, osmoadaptation does not explain Archaea that grow in low-ionic-strength habitats, such as {\it Methanobacterium thermoautotrophicum}, which has a cytosol $C_\text{K}=0.65-1.1$ M \cite{mar99}.
In cases of this type, the high $C_\text{K}$ could be a relic of ancient conditions, rather than the result of adaptation. 
If unicellular organisms close to the root of the phylogenetic tree
 preserve memory of their past history \cite{mac26,dib15}, 
the potassium-rich cytosol of Archaea is consistent with a scenario in which
life emerged in an environment with a high concentration of potassium. 
 
\subsection{Enhancement of the Radiation Activity in the Early Archean} 
\label{Archeanradioactivity}
 
Radiation sources that were enhanced in the primitive Earth might have contributed to the special
conditions that allowed life to emerge in the early Archean. 
If radiochemistry did play a role in life's origins, 
we can assess the relative biogenetic importance of different sources of natural radiation
by comparing their strength at 4 Ga and at the present time. 

Since $^{40}$K has a half-life
 of 1.248 Gyr, 
the isotopic ratio $^{40}$K/K was one order of magnitude higher
in the early Archean than today. 
Therefore, in addition to the properties discussed above,
$^{40}$K is also a potential contributor to the biogenetic conditions of the Archean. 
This is not the case for the other natural sources of radiation, as we will now discuss. 
For the reasons explained in Section \ref{compatibility}, we did not consider $\alpha$-particle emitters,
such as the heavy radionuclides shown in Table \ref{tabRadionuclides}. 

\subsubsection{Short-Lived Radionuclides of Astrophysical Origin}

Among the radionuclides of astrophysical origin,
$^{26}$Al and $^{60}$Fe decay without emitting $\alpha$ particles.
$^{26}$Al disintegrates by electron capture and $\beta^+$ emission.
$^{60}$Fe disintegrates by $\beta^-$ emission to $^{60}$Co, 
which is unstable and decays in a short time to $^{60}$Ni
through different routes with $\beta^-$ and $\gamma$ emissions.
The half-lives of these nuclides
($\tau$ = 0.71 and \mbox{2.6 Myr \cite{rug09,wal15,ost17},} for $^{26}$Al and $^{60}$Fe, respectively) 
are much shorter
than the time scale required for the Earth to become habitable. 
As a result, the $^{26}$Al and $^{60}$Fe incorporated in the Solar Nebula 
did affect the geochemical processes at the time of the formation of the Solar System (4.55 Ga \cite{lic21}),
but completely faded out at the epoch of life's origins, a few hundred Myr later. 
Even if both radionuclides could have been delivered to Earth at later epochs from explosions of nearby supernovae,
there is no reason why such explosions should have been more frequent in the Archean than at later stages. 

\subsubsection{Radiation Sources Generated by Galactic Cosmic Rays}
\label{Archeancosmicrays}

The interactions of galactic cosmic rays (GCRs) with the molecules in the highest atmospheric levels
 generate radionuclides and energetic particles that may have affected 
the atmospheric and surface chemical processes in the primitive Earth. 
Among the products of GCRs, $^{14}$C is particularly interesting because:
(i) similarly to $^{40}$K, it decays to a stable nuclear product ($^{14}$N) emitting a $\beta^-$ particle
and, (ii) given the biological role of carbon, $^{14}$C could have been incorporated 
in prebiotic molecules, 
generating internal $\beta^-$ radiation from within the molecules themselves \cite{noy77}.
However, there are reasons to believe that the Archean production rate of $^{14}$C was smaller than today.
The flux of GCRs arriving at our location in the Solar System is partially shielded by the solar wind and the solar magnetic field \cite{mar22}. 
The existence of this shielding effect is supported by the observation that the $^{14}$C production rate
is modulated by the solar cycle, being lower when the solar activity is higher \cite{jok00}.
Owing to the enhanced activity of the young Sun, the shielding must have been stronger in the Archean, when
 the GCR flux at 1 AU could have been lower than today by up to two orders of magnitude \cite{coh12}. 
As a result, the rate of $^{14}$C production must have been consistently lower at 4 Ga than today,
indicating that $^{14}$C was not a key biogenetic ingredient of the Archean.
We note that $^{14}$C can also be produced by high-energy solar particles, but only in negligible amounts \cite{kov12}.

The reduced flux of GCRs at 4 Ga also affects other products of GCRs, such as pions and muons. 
Muons are the most important contributors to the surface flux of secondary cosmic rays hitting the Earth's surface. 
They may have played a role in prebiotic chemistry and, in particular, according to a recent study \cite{glo20}, in the origin of biomolecular chirality. However, as in the case of $^{14}$C,
the enhanced activity of the young Sun must have reduced the production of muons in the Archean. 
The implications of the reduced dose of muons in the prebiotic context are discussed in Section \ref{chiralnoise}.

\subsection{Summary}

Based on its abundance and distribution on the Earth's surface, its potential role in prebiotic chemistry,
and its continuity with extant life, potassium was likely to be present in all the steps of chemical evolution that led
to the emergence of life. The same is true for its unstable isotope, $^{40}$K, which, at variance with heavy radionuclides,
generates {ionizing radiation and daughter nuclides that are not particularly harmful for protobiological molecular structures. Thanks to this fact, in a prebiotic mix with potassium compounds, the radiation emitted by
$^{40}$K can provide a moderate, but persistent effect on prebiotic molecules, potentially affecting their evolution.
Since $\beta$ particles, such as those emitted by $^{40}$K, have a much longer penetration range than $\alpha$ 
particles \cite{cho02,mag05}, $^{40}$K decays can affect a relatively large volume of a prebiotic solution.} 
At variance with other natural radionuclides that emit $\beta$ particles, 
the isotopic abundance of $^{40}$K was enhanced in the Archean. 
For all the above reasons, 
$^{40}$K is the most likely radiation source that may have influenced the advanced stages
of prebiotic molecular evolution. The possibility that 
$^{40}$K might have triggered the chirality of biomolecules is discussed in Section \ref{chirality}.

\section{Radiochemical Impact of $^{40}$K in the Primitive Earth} 
\label{40Kradiochemistry}

Most $^{40}$K nuclides decay through $\beta^-$ decay, which takes place in 89.3\% of cases, 
or electron capture followed by a $\gamma$ ray emission, 
which takes place in 10.7\% of cases. 
To estimate the radiochemical impact in prebiotic chemistry, here, we restrict our attention
to the $\beta^-$ decay route, not only because it
takes place more frequently, but also because it gives a larger linear energy transfer (LET) 
and provides chiral effects (Appendix \ref{properties40K}).
We considered a mix of simple organic molecules and potassium compounds dissolved in a solution and
 assumed that the solvent is composed of water and/or formamide \cite{vla18,sal19}.


\subsection{Radiation Dose of $\beta^-$ Particles}
\label{beta40Kimpact}

A number $N_\circ$ of unstable nuclides with half-life $\tau$ decrease with time according to the law 
$N(t)= N_\circ \, \text{e} ^{ - \ln2 \, (t/\tau) }$ \cite{cho02}. 
Therefore, the isotopic ratio $r_\text{40} \equiv ^{40}$K/K 
 increases with time before the present, $t_\text{BP}$, according to the expression:
\begin{equation}
r_\text{40}(t_\text{BP})=r_\text{40}(0) \, e^{ \, \ln2 \, ( t_\text{BP}/\tau_{40\text{K}} ) }~~,
\label{timeEvolution}
\end{equation}
where $r_\text{40}(0) = 1.17 \times 10^{-4}$ is the present-day ratio
and $\tau_{40\text{K}}=3.94 \times 10^{16} \, \text{s}$ (Table \ref{tabRadionuclides}). 
The activity generated by an ensemble of $N$ unstable nuclides 
with half-life $\tau$ is $A_i= p_i \, (\ln 2/\tau) \, N$ (Bq), 
where $p_i$ is the probability of the $i$-th decay mode \cite{cho02}. 
For the $\beta^-$ decay mode of radioactive potassium $p_i=0.893$,
the decay rate per unit volume (Bq/L) is
\begin{equation}
a_{\beta} (t_\text{BP}) = 0.893 \, { \ln 2 \over \tau_{40\text{K}} } \, r_\text{40} (t_\text{BP})\, N_A C_\text{K} ~~,
\label{Activity}
\end{equation}
where 
$C_\text{K}$ is the molar concentration of K$^+$ ions in the solution (M = mole/liter) and
$N_A$ is the Avogadro number. 
Rates of $\beta^-$ decays per unit volume calculated with \eqref{Activity} 
for $C_\text{K}=0.25,0.5$ and 1\,M are shown in Table \ref{tabActivity}.
These three values of potassium concentration correspond to $5\%$, $10\%$, and $20\%$, respectively, 
of the saturation level of KCl in water ($C_K=5$\,M at $T=25\,^\circ$C). 
High concentrations of K$^+$ ions can also be obtained 
by dissolving 
compounds that are expected to be present 
in evaporating alkaline lakes considered in studies of life's origins \cite{ton19}. 
Values as high as $C_K=1$ \,M
have been used in successful experiments of assemblage of peptides \cite{dub13}.
Values in the range $C_K \simeq 0.5$ to $1$\,M are representative of the concentration 
in the cytosol of many Archaea \cite{mar99}. 
\begin{table}[H]
\caption{Rate 
 of $\beta^-$ decays per unit volume in a solution of potassium salts. 
\label{tabActivity}}
\newcolumntype{C}{>{\centering\arraybackslash}X}
\begin{tabularx}{\textwidth}{CCCC}
\toprule
 & \boldmath{$C_K=0.25\,\mathrm{M}$} & \boldmath{$C_K=0.5\,\mathrm{M}$}& \boldmath{$C_K=1.0\,\mathrm{M}$} \\ 
 \midrule
\boldmath{$t_\text{\textbf{BP}}$} & \boldmath{$a_{\beta}$} & \boldmath{$a_{\beta}$} & \boldmath{$a_{\beta}$} \\ 
\textbf{(Ga)} & \boldmath{\textbf{(Bq L}$^{-1}$\textbf{)}} & \boldmath{\textbf{(Bq L}$^{-1}$\textbf{)}} & \boldmath{\textbf{(Bq L}$^{-1}$\textbf{)}}\\
\midrule
0 & $ 2.77 \times 10^2$ & $5.54 \times 10^2$ & $1.11 \times 10^3$ \\ 
4 & $ 2.55 \times 10^3$ & $5.11 \times 10^3$ & $1.02 \times 10^4$ \\ 
\bottomrule
\end{tabularx}
\end{table}


For the sake of comparison with other natural radiation sources,
the activity per unit volume calculated with \eqref{Activity} can be converted to an annual equivalent dose (Sv/yr): 
\begin{equation}
d_{\beta}(t_\text{BP}) = 3.16 \times 10^7 \, E_{\beta} \, a_{\beta} (t_\text{BP}) \, \rho^{-1} ~~,
\label{dose40}
\end{equation}
where $E_{\beta} =0.499 \,\text{MeV} =8.0 \times 10^{14}$\,J is the mean energy of the $\beta^-$ particle, 
$\rho$ is the mean density (kg/liter) of the solution, and the constant converts seconds into years.
A radiation weighting factor $W_{\beta} \equiv 1$ has been adopted for the $\beta^-$ particles \cite{cho02}.
We take $\rho \simeq 1$ kg/dm$^3$ as the density of water;
this value is also appropriate for a mix of water and formamide,
because $\rho \simeq 1.13$ kg/dm$^3$ for formamide.
%

\begin{figure}[H]  

\includegraphics[width=8 cm]{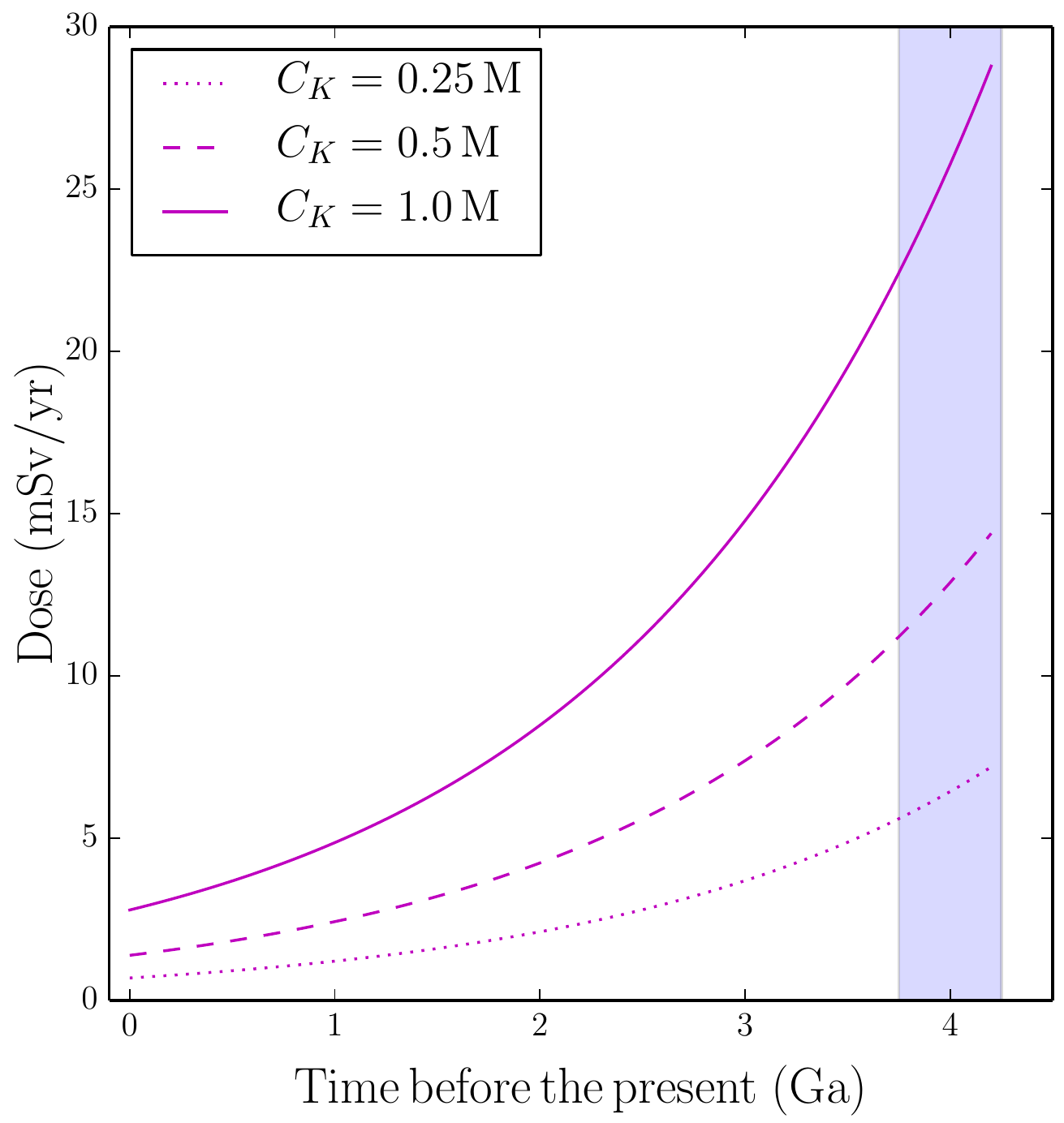} 
\caption{
Dose of $\beta^-$ radiation calculated with Equation \eqref{dose40} for
aqueous solutions of potassium ions with molar concentration, $C_K$, indicated in the legend.
The evolution of the $^{40}$K/K isotopic ratio from the time of Earth's formation to the present time
is calculated with Equation \eqref{timeEvolution}.
The vertical band indicates the interval during which life emerged on Earth.
}
\label{figA40K}%
\end{figure}

\textls[-15]{The evolution of $d_{\beta}(t_\text{BP})$ at constant $C_\text{K}=0.25,0.5$ and 1\,M,
calculated with \mbox{Equation \eqref{dose40},}} is shown in Figure \ref{figA40K}.
The dose at $t_\text{BP}\simeq 4$\,Ga is $d_{\beta}=$ 6.4, 12.9, and 25.8 mSv/yr
for \mbox{$C_\text{K}$= 0.25,} 0.5, and 1\,M, respectively.
These values can be compared with the present-day doses of the Earth radiation background
from the ground (0.48 mSv/yr) and cosmic rays \mbox{(0.39 mSv/yr) \cite{UN10}.}
These estimates indicate that 
the dose of $\beta^-$ particles could have been one or two orders of magnitude larger in the prebiotic world than today. 
 
%


High values of concentration can be attained in the framework of dry--wet scenarios 
of the origins of life in subaerial environments \cite{mul12,ton19,dam20}. 
The fact that the radiation dose depends on the concentration of the solution
means that the impact of $^{40}$K decays is stronger 
when the solution is less hydrated and, so, more favorable to condensation.
In turn, this implies that $\beta^-$ particles may influence 
the structural (and, hence, functional) properties of early polymers
at the very moment of their synthesis.

\subsection{Indirect Effects on Prebiotic Molecules Dissolved in Diluted Solutions} 
\label{decaysvol}

In diluted solutions, the primary events generated by the $\beta^-$ particles 
impact the solvent, rather than the solutes.
The organic molecules dissolved in the solution are affected by the 
secondary events triggered by the primary events. 
If $V_t$ is the total volume of the solution and $V_s$ the volume influenced by a chain of secondary events, 
the probability that one point of the solution is reached by these events is $\approx V_s / V_t$. 
Multiplying this probability by the rate of primary events, $a_{\beta} \, V_t$,
we obtain the rate of secondary events that affect one specific point of the solution:
 \begin{equation}
 A_s \approx a_{\beta} \, V_s ~. 
 \label{ratesecondaryhits}
 \end{equation}
For randomly distributed prebiotic molecules, $A_s$ represents 
the fraction of prebiotic molecules 
involved in a secondary event per unit time. 
Equation \eqref{ratesecondaryhits} is correct if the volumes $V_s$ generated by different primary events do not overlap; 
this is a reasonable approximation for low doses of radiation, as in our case.

If we call $C_m$ the molar concentration of the prebiotic molecules in the solution, 
the total number of prebiotic molecules in the mix is $N_A \, C_m \, V_t$, and
 the number of prebiotic molecules that are reached by a secondary event per unit time is
\begin{equation}
 A_{m} \approx A_s \, N_A \, C_m \, V_t~. 
\label{eqAm}
\end{equation} 
To estimate $A_m$, we considered a prebiotic solution at $t_\text{BP}=4$\,Ga with $C_\text{K}=1$\,M.
We took $V_t$ equal to the volume of a sphere of radius 1\,$\mu$m
(the typical size of a prokaryotic cell), which we adopted here as representative of a protocell embedded in the solution. 
To calculate $A_s$, we considered two models for the propagation of secondary events.

\subsubsection{String of Spurs Model}
\label{spurs}

The primary events are distributed as a string of beads, called spurs, along the path of the $\beta^-$ particle \cite{dra93}.
The secondary events propagate inside the spurs, which we modeled as spheres of radius $r_s$. 
The approximate number of spurs generated by a primary event is $\approx$$x_p / \ell_s$,
where $x_p$ is the penetration range of the $\beta^-$ particles and $\ell_s$ the distance between spurs.
In this schematic model, 
the volume of solution where prebiotic molecules can be affected is $V_s \approx (x_p / \ell_s) \, (4/3) \pi \, r_s^3$.
To estimate $V_s$, we adopted $x_p \simeq 4.1$\,mm, the penetration range of $\beta^-$ particles in water \cite{cho02}, 
and $\ell_s \simeq 0.16 \mu$m, the typical distance between ion pairs formed along the track of the beta particle in water
(Appendix \ref{betaminusroute}). 
The effective radius $r_s$ depends on the capability of a secondary event to propagate effects through the solvent.

\begin{figure}[H]  
\includegraphics[width=8 cm]{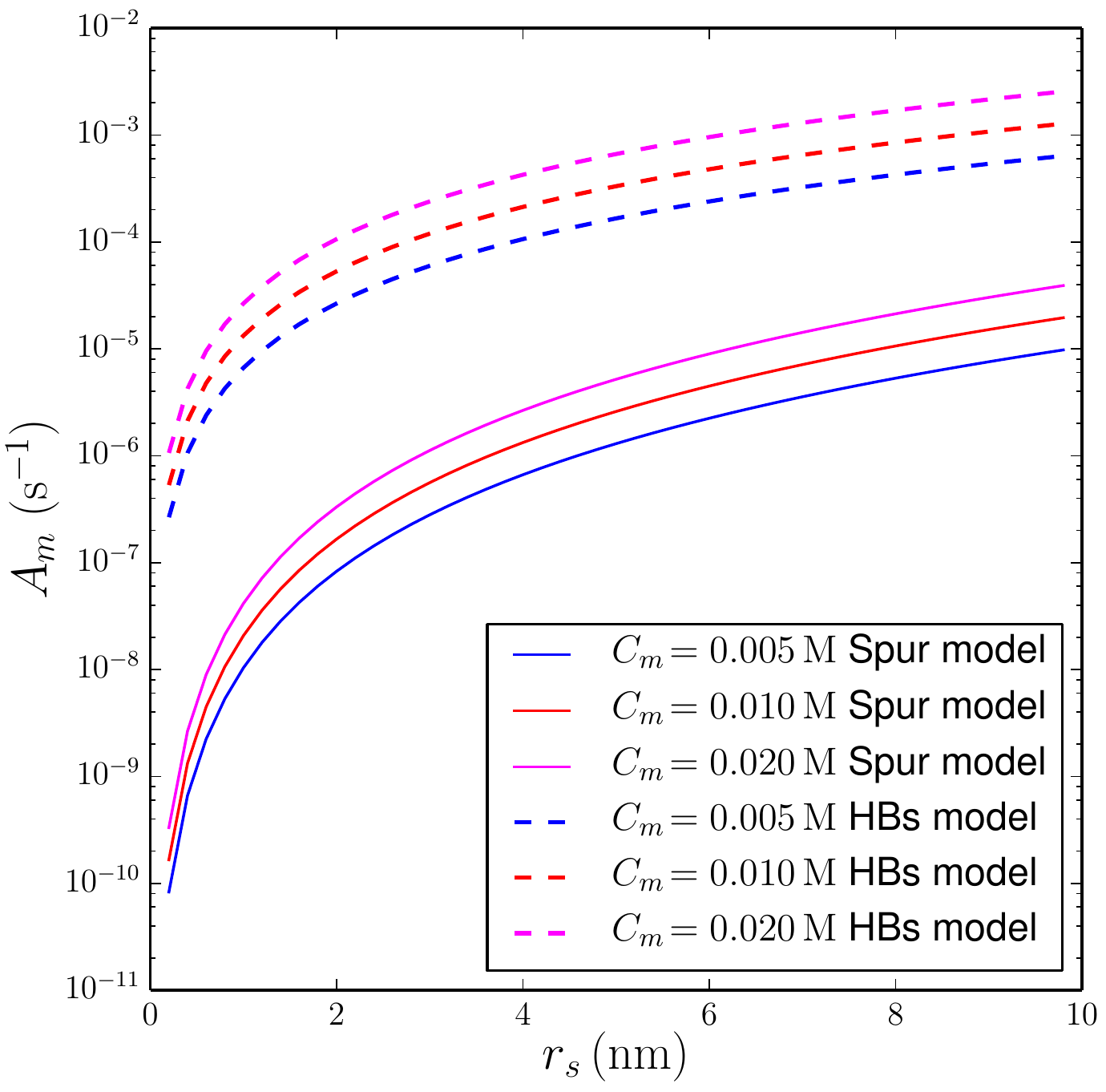} 
\caption{
Effects of $\beta^-$ decays of $^{40}$K on prebiotic molecules dissolved in an aqueous solution with potassium ions ($C_K=1$\,M)
at $t_\text{BP}=4$\,Ga.
Vertical axis: number of prebiotic molecules affected by secondary events per unit time 
within a sphere of radius 1\,$\upmu$m (representative size of a protocell). 
Horizontal axis: effective radius of influence of secondary events, $r_s$. 
Each curve was obtained at a constant molar concentration of prebiotic molecules, $C_m$, indicated in the legend.
Solid lines: string of spurs model of the propagation of secondary events (Section \ref{spurs}).
Dashed lines: model of the propagation of secondary events via hydrogen bonding (Section \ref{HBprop}). }
\label{tsecondaryhits}%
\end{figure}

Examples of calculations of $ A_{m}$ versus $r_s$ with the string of spurs model
are shown as solid curves in Figure \ref{tsecondaryhits}, 
for three values of $C_m$ in the millimolar range, indicated in the legend.
For a typical spur radius,
$r_s = 1$ nm and 
$C_m \simeq 10$ mM, the number of prebiotic molecules affected by a secondary event per unit time is 
$ A_{m} \approx 2 \times 10^{-8}$ s$^{-1}$, corresponding to an event every $\simeq 1.5$ yr for a given molecule.
The predicted frequency of interactions becomes larger increasing the concentration of prebiotic molecules
and/or the effective radius of impact of secondary events. 
For $r_s = 3$ nm (about 10-times the intermolecular distance of water molecules) and $C_m \simeq 20$ mM,
we obtain $ A_{m} \approx 1 \times 10^{-6}$ s$^{-1}$, i.e., an event every $\simeq 10$ days. 
Since the spur radius is probably closer to $\approx$1 nm,
the above results indicate that, in order to impact a significant fraction of prebiotic molecules,
the effects of secondary events should accumulate over long periods of time.
 

%

\subsubsection{Propagation of Secondary Effects via Hydrogen Bonding}
\label{HBprop}

The average LET of the $\beta^-$ decays of $^{40}$K implies that
the mean energy transferred by the fast electron to each water molecule is $\simeq$0.07 eV
(Appendix \ref{betaminusroute}).
This value is in the range of dissociation energies of hydrogen bonds. 
Therefore, besides the effects that create the spurs
(excitation and ionization of molecules along the path),
we expect that the hydrogen bonds of the solvent will be rearranged
along the whole path of the $\beta^-$ particle. 
Hydrogen bonds are well known
for their cooperative behavior \cite{DS99,Marechal07,Gilli09,vla18}. 
%
Owing to this property, each rearrangement of hydrogen bonds 
may become a new source
 of secondary effects. For instance, since hydrogen bonds have an electrostatic component, the cooperative behavior will affect 
the polarization state of the prebiotic molecules involved in these secondary events. 
This might affect prebiotic molecular evolution because 
the polarization state plays a key role in governing the structure and functionality of organic molecules. 
Therefore, the propagation of the indirect effects of the $\beta^-$ particles might influence
dissolved molecules over a distance range much larger than the mean intermolecular distance of water molecules ($\approx$0.3 nm). 
This could happen, for instance, if the orientation of the water molecules undergoes a transition
due to the electric polarization induced by the passage of the fast, negatively charged electron. 

Based on the above arguments, we re-calculated the rate of secondary impacts of the prebiotic molecules
using a schematic geometrical configuration different from the string of spurs model. 
In practice, to estimate the volume $V_s$ in Equation \eqref{ratesecondaryhits},
we assumed that: (i) the $\beta^-$ particle influences the solvent along its whole path (equal to the penetration range, $x_p$),
and (ii) at each point of the path, the cooperative effects of hydrogen bonding propagate the secondary effects 
in circles of effective radius $r_s$. In this idealized scheme, $V_s \approx x_p \, (4 \pi r_s^2)$.
Estimates obtained with this schematic model are shown as dashed lines in Figure \ref{tsecondaryhits}.
One can see that, for $r_s = 1$ nm and $C_m \simeq 10$ mM, 
the number of prebiotic molecules affected by a secondary event per unit time is 
$ A_{m} \approx 1.3 \times 10^{-5}$ s$^{-1}$, corresponding to an event \mbox{every $\simeq 21$} h for a given molecule. 
If secondary effects propagate for $r_s = 3$ nm and \mbox{$C_m \simeq 20$ mM,} 
we obtain $ A_{m} \approx 2.4 \times 10^{-4}$ s$^{-1}$, corresponding to an event per hour. 
Assuming that the cooperative properties of hydrogen bonding are able to propagate 
the indirect effects over a distance of a few tens of water molecules, the rate would be even stronger.
For $r_s = 10$ nm and $C_m \simeq 10$ mM, 
we obtain $ A_{m} \approx 1.3 \times 10^{-3}$ s$^{-1}$, i.e., an event every 12 minutes. 
Time scales of this order of magnitude could be sufficiently fast to influence the evolution of prebiotic molecules in scenarios
of the origin of life in aqueous ponds or lakes. 

Understanding if the hydrogen bond model of propagation is plausible 
and estimating its radius of action are critical to assess the potential impact of $^{40}$K decays in a diluted solution. 
The effective range of influence of secondary effects via hydrogen bonding could be tested with the aid of
specific experiments or molecular simulations. 
This model of propagation is relevant to any solvent capable of forming a network of hydrogen bonds. 
Apart from water, an example is formamide, a prebiotic solvent that is able to form a network of hydrogen bonds
at least as efficient as that of water \cite{vla18}. 

\subsection{Non-Diluted Solutions}

In the dry phase of dry--wet scenarios of the origins of life, the concentration of solutes is very high,
and the rates presented above, calculated for diluted solutions, are not valid anymore.
The same is true for an intermediate moist phase, with characteristics of a hydrogel, which is expected to exist 
in the dry--wet scenarios \cite{dam20}. 
Treating the propagation of $\beta^-$ particles in dry or hydrogel phases is beyond the purpose of this work. 
We expect, however, that the high concentration of potassium ions in these phases would yield a
higher radiation dose, which, together with a higher concentration of prebiotic molecules,
would result in a higher frequency of events per molecule.

\section{Chiral Effects}
\label{chirality}

The direct and indirect effects of $^{40}$K decays discussed
in the previous section may have influenced the prebiotic evolution of molecular structures in different ways. 
Here, we consider the potential impact on molecular chirality. 
Parity violation of electroweak forces is a natural process, which, in principle,
may provide an enantiomeric excess with a well-defined handedness \cite{lee56,wu57,kes77,sal91}. 
Two types of electroweak forces can provide a chiral asymmetry:
(i) weak neutral currents, mediated by a $Z^0$ boson, and
(ii) weak charged currents, mediated by a $W^+$ or $W^-$ boson.
The chiral signal expected by weak neutral currents is too small to be practically detectable
 \cite{mac89,fag01,fag05}. 
Theoretical calculations indicate that the chiral asymmetry induced by weakly charged currents,
such as those involved in the $\beta$ decay (Appendix \ref{betaminusroute}), 
should be orders of magnitude larger \cite{heg85b,heg87}. 
Therefore, the $\beta$ decay is the best candidate for providing an enantiomeric excess 
due to parity violation of weak nuclear forces. 

Several searches for enantiomeric excess 
induced by $\beta^-$ or $\beta^+$ sources or artificially polarized electrons
have been performed in the past. 
Experiments with natural $\beta^-$ decay sources found differential decomposition 
favoring the survival of L-type amino acids in some cases \cite{gar68,dar76}, 
but no differential decomposition in another case \cite{bon79}.
Evidence for left-handed, circularly polarized bremsstrahlung radiation was also detected \cite{gol57}.
Tests performed with natural $\beta^+$ decay sources found differences in annihilation intensity favoring
the destruction of D-type amino acids \cite{gar73,gar74}. Experiments with longitudinally polarized electrons showed that
left-handed electrons degrade D-leucine more efficiently than right-handed electrons \cite{bon75},
but only an upper limit on molecular asymmetry was found using positrons with positive helicity \cite{gid82}.

All together, the experimental evidence suggests that $\beta^-$ decays may trigger a small 
enantiomeric excess of L-type amino acids. 
Among the sources discussed here, two candidate $\beta^-$ emitters are $^{40}$K or $^{14}$C,
but the $\beta^-$ particles emitted by $^{40}$K have a higher energy and, therefore, a higher helicity (\cite{mei87} Figure 3)
than those emitted by $^{14}$C. 
Therefore, $^{40}$K is a better candidate for the origin of biomolecular chirality than $^{14}$C, 
not only because it was more enhanced in the Archean than today (Section 
\ref{Archeancosmicrays}), but also because of the higher intensity of its chiral signal.

To be sure that $\beta^-$ decays can indeed trigger an enantiomeric excess, we should
try to understand why the experiments do not provide clear-cut results.
A possible explanation is that the experiments did not approach the level of sensitivity necessary to observe the small asymmetries predicted by theoretical calculations \cite{gid82}. 
Apart from this, there are other reasons that may explain why it is difficult to detect the chiral signal. 
In the next two sub-sections, we present two possible reasons and discuss the implications 
in the context of Archean scenarios of abiogenesis. 

\subsection{Chiral Noise}
\label{chiralnoise}

A possible source of the disturbance of the chiral effects generated by $\beta^-$ decays is the presence of particles, which, having opposite helicities, introduce a sort of ``chiral noise''. This is the case of the muons generated
in air showers, which are the most relevant by-products of GCRs that hit the ground.
Positive and negative muons are generated in the high atmosphere by the decays of positive pions:
\begin{equation}
\pi^+ \rightarrow \mu^+ + \nu_\mu ~,
\label{piplusdecay}
\end{equation}
and negative pions:
\begin{equation}
\pi^- \rightarrow \mu^- + \overline{\nu}_\mu ~,
\label{piminusdecay}
\end{equation} 
respectively.
Since pions have zero spin and their decays are governed by weak nuclear forces, 
the opposite outgoing leptons must have the same helicity state \cite{mar17}. 
In practice, the $\mu^+$ generated in the decay \eqref{piplusdecay} has always negative helicity, whereas
the $\mu^-$ generated in the decay \eqref{piminusdecay} has always positive helicity. 
Therefore, the muons will induce opposite chiral effects on the molecules of a prebiotic solution with potassium ions.
The possibility for $^{40}$K to induce a chiral effect in the solution without a significant chiral noise
will depend on the relative dose of muons and $\beta^-$ particles.
The present-day radiation dose of muons hitting the ground is 0.13 mSv/yr \cite{atr13}. 
This value is probably a stringent upper limit of the Archean dose, since the GCRs arriving at 1 AU
were efficiently shielded by the enhanced solar wind of the young Sun. 
The dose of $\beta^-$ in an Archean solution with potassium ions
is >6.4 mSv/yr for $C_\text{K} > 0.25$\,M (Section \ref{beta40Kimpact}). 
Therefore, we do not expect that muons in the primitive Earth were able to disturb
the chiral effects induced by $^{40}$K, even in a prebiotic soup with a moderate concentration of potassium. 
The Archean conditions might have been favorable to the generation of a small
enantiomeric excess thanks to a lower flux of muons (see, however, \cite{glo20}).

\subsection{Propagation of Chiral Effects in the Solvent}

A problem of the propagation of chiral effects is that compounds dissolved in diluted solutions 
are not affected directly by the chiral particles, but react in most cases with the solvent and its radiolysis products.
The free radicals generated by radiolysis have single electrons with non-zero spin, which, in principle,
could propagate the helicity of the $\beta^-$ particle. 
%
However, 
chiral effects can hardly propagate in a non-chiral solvent such as \mbox{water \cite{zag93}.}
To bypass this problem in the context of life's origins, we can envisage a prebiotic scenario where organic oligomers are dissolved in a chiral solvent. 
In this scenario, the chiral molecules of the solvent might be able to propagate the chiral effects.
A very effective solvent in the context of prebiotic studies is formamide, 
which is able to assist a broad spectrum of synthesis reactions \cite{sal19}
and creates a pervasive network of hydrogen \mbox{bonds \cite{vla18}. }
The formamide molecule is achiral in its ground state, but can become chiral by populating its excited states with some external energy source,
such as near UV \mbox{photons \cite{rou17}.} In a prebiotic soup with potassium ions,
formamide could be {frequently excited to its chiral state 
by the abundant flux of UV photons} hitting the Earth's surface in the Archean. 
Therefore, {assuming the existence of prebiotic environments with high concentrations of formamide},
the possibility of propagating chiral effects through a solvent in a prebiotic soup is plausible, 
but should be tested in dedicated experiments. 

\subsection{Simulating Chiral Effects in Archean Conditions}

Based on the above discussion, one can envisage the following experiment to mimic a scenario of chiral effects
naturally induced in a surface pond with prebiotic molecules and potassium compounds dissolved in formamide
(or water plus formamide): 
(i) the formamide molecules are excited to their chiral states by 
the $\beta$ decays or by the UV radiation impinging on the pond; 
(ii) the chiral states of formamide are influenced by the helicity of the $\beta^-$ particles emitted by $^{40}$K;
(iii) the chiral formamide acts as an intermediate product able to propagate the chirality of the $\beta^-$ particles
to the prebiotic molecules embedded in the soup. Experiments of this type, using potassium salts
with enriched $^{40}$K/K isotopic ratio representative of the Archean, should be performed to understand
if the conditions of the primitive Earth were conducive to the generation of an initial enantiomeric excess.
To the best of our knowledge,
the impact of chiral $\beta^-$ particles constantly injected in a (prebiotic) molecular medium has not been investigated so far.
Some screening of present-day muons could be performed to simulate the shielding effect from GCRs resulting from the enhanced activity of the young Sun.

\section{Conclusions}
\label{conclusions}

The abundance and distribution of chemical elements on the Earth's crust and in the sea and the continuity with the properties of extant life suggest that potassium, in its stable and unstable isotopic forms, was likely present in all stages of terrestrial biogenesis,
from the abiotic assemblage of monomers--oligomers, up to the emergence of protocells. 
At variance with the U and Th radionuclides, which emit $\alpha$ particles
and unstable nuclei, $^{40}$K decays to stable nuclei emitting $\beta^-$ and $\gamma$ rays, which do not destroy the covalent bond structure
of early biomolecules and protocells. At variance with other mild sources of natural radiation, 
$^{40}$K was more abundant in the Archean than today 
and may have contributed to the special, perhaps unique, biogenetic conditions of the primitive Earth.

%
%
%
%
%
%

In the most-frequent $\beta$-decay route of $^{40}$K, 
an Archean prebiotic solution with plausible values of the concentration of potassium ions,
$C_\text{K}=0.25$ to 1.0\,M, generates an effective radiation dose
$d_{\beta}=$ 6.4 to 25.8 mSv/yr, one to two orders magnitude larger than the present-day radiation background on Earth. 
%
%
The dependence of the dose on the concentration and the possibility 
that $^{40}$K has an active biogenetic role
are particularly interesting in the framework of dry--wet scenarios of life's origins: 
the dose is stronger when the solution is less hydrated and more favorable to dehydration synthesis.
Therefore, the internal radiation of $^{40}$K
might influence the structural and functional properties of early polymers
at the very time of their assemblage. 

The rates of $\beta^-$ events affecting organic molecules dissolved in a solution with K ions
depend on the efficiency of the propagation of secondary effects through the solvent.
To estimate such rates, we considered two models of propagation in diluted solutions:
(i) a spur model that accounts for the radiolysis products generated by the primary events;
(ii) a model based on cooperative effects of the hydrogen bond network of the solvent.
%
In the spur model, given a concentration of organic molecules of $\simeq$10 mM, 
the rates inside a representative volume of a protocell
are low ($\approx$$2 \times 10^8$\,s$^{-1}$ for $C_\text{K}=1.0$\,M), and long times of integration are required
to accumulate a significant radiogenic impact. 
The rates are $\gtrapprox$500-times higher in the hydrogen bond model of propagation,
but the validity of this model should be tested with molecular simulations or experiments. 
Higher rates of impacts on prebiotic molecules are expected in non-diluted solutions, such as the dry or hydrogel phases 
of dry--wet scenarios of the origins of life. 

The left-handed $\beta^-$ particles emitted by 
the $^{40}$K decay
tend to generate a small enantiomeric excess of L-type amino acids. 
The chiral signal is expected to be very small,
but the persistent injection of internal $\beta^-$ particles may lead to an accumulation of the effect
in protobiological molecular structures. 
In the Archean scenario of life's origins, $\beta^-$ particles might have been
able to generate a chiral imprint better than today because: (i) the flux of secondary muons, which
create a sort of chiral noise, was reduced due to the enhanced shield from cosmic rays of the young solar wind;
(ii) a commonly invoked prebiotic solvent, formamide, might have been more effective to propagate chiral effects than water. 

The results of this work suggest that $^{40}$K should be included in realistic prebiotic experiments. 
Recipes for prebiotic soups that simulate Archean conditions {may consider some of} the following ingredients:
(i) potassium compounds with an enriched $^{40}$K/K isotopic ratio; (ii) a mix of water and formamide as a solvent; (ii) irradiation by UV photons; (iii) partial screening of cosmic muons.

\vspace{6pt} 



\vspace{+6pt}

\funding{This research was funded by the Italian Space Agency (Life in Space project, ASI N. 2019-3-U.0).}
\institutionalreview{Not applicable}
\informedconsent{Not applicable}
\dataavailability{Not applicable} 

\acknowledgments{I wish to thank Raffaele Saladino and Ernesto Di Mauro for their useful comments
on an early version of this manuscript {and Piero Ugliengo for a fruitful discussion on several topics
raised in this paper. The suggestions of an anonymous Reviewer were important
to clarify some aspects of the radiochemical impact of $^{40}$K.}}

\conflictsofinterest{The author declares no conflict of interest.} 



%

\appendixtitles{yes} 
\appendixstart
\appendix

\section[\appendixname~\thesection]{Basic Properties of \boldmath{$^{40}$}K Decay}
\label{properties40K}

\subsection[\appendixname~\thesubsection]{The $\beta^-$ Decay Route} 
\label{betaminusroute}

In the $\beta^-$ decay mode, which takes place in 89.3\% of cases, a neutron of the $^{40}$K nucleus is converted 
into a proton with the emission of a $\beta^-$ particle and an 
antineutrino of the electron type: 
\begin{equation}
^{40}_{19}\text{K} \rightarrow [^{40}_{20}\text{Ca}]^+ + \beta^- + \overline{\nu}_e ~~.
\label{route1}
\end{equation}
The decay is mediated by a $W^-$ boson and is classified as a weakly ``charged current'' interaction
because the charge of the initial and final fermion (a nucleon in this case) \mbox{changes (\cite{bet14}}, Chapter 7).
The antineutrino $\overline{\nu}_e$ escapes without interacting with the medium,
and the radiochemical impact is provided by the $\beta^-$ particle and the recoiling nucleus, $[^{40}_{20}\text{Ca}]^+$. 
The total emitted energy is constant, but is distributed between the $\beta^-$ particle and the antineutrino.
Depending on the energy taken by the antineutrino, the fast electron is emitted with an energy that ranges from zero to 
the end-point energy of 1.311 MeV,
with a mean value of $0.499$ MeV \cite{cam04}. 
At these energies and for a prebiotic medium with a low atomic number, 
the dominant routes of energy loss of the $\beta^-$ particle are ionization and excitation
of the atoms/molecules in the medium, 
each process contributing approximately half \cite{cho02}.
For each ionization or excitation, the particle is deflected with a wide angle, leading to an erratic path;
secondary ionizations can be produced along the whole path \cite{cho02}. 
The large number of interactions and the erratic path imply a short range of penetration and
a large value of linear energy transfer (LET), i.e., 
 the energy absorbed in matter per unit path length traveled by a charged particle \cite{cho02}.
The maximum range of penetration of a 1~MeV electron in water is $4.1$ mm, 
with an average LET
value of 0.24 keV/$\upmu$m (\cite{cho02} Table~6.2).
Considering that the average energy for the formation of ion pairs in water is \mbox{38\,eV (\cite{cho02}} Table 7.1),
the average LET value generates half a dozen ion pairs/$\upmu$m. 
Therefore, the typical distance between ion pairs 
is $\approx$$0.16 \, \upmu$m. 
For a typical intermolecular separation of 3\,\AA\ between water molecules,
the mean energy transferred by the fast electron to each water molecule is $\simeq$0.07 eV. 

%
{{\it Chiral effects:}} 
 The $\beta$ decay is a product of weak interactions, which 
 are characterized by parity violation \cite{bet14}. 
As a result, the $\beta$ decay generates
an asymmetry in the helicity of the emitted particles, i.e., an asymmetry in the 
spin component of the particles in their direction of motion.
%
If $\overline{p}$ is the momentum of the particle and $\overline{\sigma}$ its spin,
the helicity operator is ${1 \over 2} { \overline{p} \cdot \overline{\sigma} \over p}.$
The two helicity eigenvalues are +1/2, if the spin is in the direction of the motion and $-1/2$ if in the opposite direction \cite{bet14}.
The states of positive and negative helicity are called right-handed and left-handed, respectively,
since the spin direction corresponds to rotational motion in the right-handed or left-handed sense
when viewed along the momentum direction \cite{mar17}. 
As a result of the decay route \eqref{route1}, 
the antineutrino $\overline{\nu}_e$, with positive helicity, does not interact with the medium, 
but the $\beta^-$ particle, with negative helicity,
may couple its spin with that of the molecules encountered along its path.
This might force the chirality of the intervening prebiotic molecules (see Section \ref{chirality}). 

{\it Bremsstrahlung radiation}:
A secondary effect of the $\beta^-$ decay is the Bremsstrahlung radiation resulting from the braking of the $\beta^-$ particle.
The slowing down is due to the interaction with the electric fields 
of the surrounding atoms' medium
and generates high-energy, highly penetrating radiation \cite{mag05} (p. 66). 
When the polarized electrons emitted in the beta decay 
slow down, they lose some of their energy by emitting left circularly polarized bremsstrahlung radiation \cite{lee56,gol57}.
Furthermore, this radiation might affect the isomerism of intervening molecules. 

{\it Effects of the daughter nucleus:} 
The ionized daughter nucleus $[^{40}_{20}\text{Ca}]^+$ is a free radical that will affect the nearby molecular constituents
through chemical interactions and recoil energy.
The recoil energy is maximum when both particles are emitted in the same direction 
or if all the energy is carried away with one of the particles.
The maximum recoil energy of $[^{40}_{20}\text{Ca}]^+$ is 41 eV (\cite{cho02} Equation (4.32)).
The mean value, in the order of $\approx 20$\,eV, is several times the dissociation energy of covalent bonds
and may cause structural rearrangements or excitation/ionization in the nearby molecules.
These effects are local because the penetration range of this ion is expected to be
 shorter than that of the $\beta^-$ particle. 



\subsection[\appendixname~\thesubsection]{The Electron Capture Route} 
\label{ECroute}

The $^{40}$K nuclei decay via electron capture, 
in 10.7\% of cases. 
In the electron capture mode, an inner electron of the atom is captured by a proton of the nucleus 
via weak, charged current interaction mediated by a $W^+$ boson \cite{bet14}.
As a result, both the nucleus and the electronic structure are changed.
In the case of $^{40}$K, this process generates the transformation:
\begin{equation}
^{40}_{19}\text{K} \rightarrow \, ^{40}_{18}\text{Ar}^* + \nu_e ~~,
\label{route2}
\end{equation}
where the daughter nuclide, $^{40}_{18}\text{Ar}^*$, is in an excited nuclear state that
decays to the ground state with the emission of a $\gamma$ ray
after an extremely short time (<10$^{-9}$\,s). 
Meanwhile, an outer electron replaces the electron that was captured, leading to 
a rearrangement of the electronic orbitals, which, in some cases, can generate 
very-low-energy Auger electrons (in the Auger effect, an electron is ejected from the atom's electron shell due to interactions between the atom's electrons in the process of seeking a lower-energy electron state).
The emitted neutrino $\nu_e$ carries the entire decay energy without interacting with the medium.
The impact on the medium is provided by the $\gamma$ photon, the Auger electrons,
and the recoil of the daughter nucleus. 

{
{\it Effects of the $\gamma$ photons:} 
The $\gamma$ photon emitted by the de-excitement of the $^{40}_{18}\text{Ar}^*$
nuclide has an energy of 1.5 MeV. 
In this energy range, the dominant mode of interaction of $\gamma$ rays 
is Compton scattering, which leads to the emission of weakly bound \mbox{electrons \cite{cho02}.}
The scattered $\gamma$ rays are highly penetrating and escape from the medium,
while the Compton electrons are braked inside the medium. 
All together, the LET is not very efficient because a large fraction of the energy 
is carried away by the $\gamma$ photon.
Therefore, most of the radiochemical impact resulting from electron capture
is provided indirectly by the Compton electrons.
However, the overall energy transmitted by these electrons is smaller than that provided by the $\beta^-$ particles,
and at variance with the $\beta^-$ particles, the Compton electrons do not have helicity. 
}
%
 
\begin{adjustwidth}{-\extralength}{0cm}

\reftitle{References}

\end{adjustwidth}
\end{document}